\documentclass[reprint, 10pt,
prx,
aps, 
amsmath,amssymb,
superscriptaddress,
floatfix
]{revtex4-2}

\usepackage{circuitikz}
\usepackage{siunitx}
\usepackage{multirow}
\usepackage{comment}
\usepackage{braket}

\begin{document}
\title{A proposal for charge basis tomography of superconducting qubits}

\author{Elena Lupo}
\affiliation{Advanced Technology Institute and School of Mathematics and Physics, University of Surrey, Guildford, GU2 7XH, UK}
\affiliation{Quantum Computing Analytics (PGI-12),
Forschungszentrum J\"{u}lich, 52425 J\"{u}lich, Germany}

\author{Daniel Long}
\affiliation{Blackett Laboratory, Imperial College London, South Kensington Campus, London SW7 2AZ, United Kingdom}

\author{Daniel Dahan}
\affiliation{Department of Physics, Ben-Gurion University of the Negev, Beer-Sheva 84105, Israel}

\author{Konstantin Yavilberg}
\affiliation{Department of Physics, Ben-Gurion University of the Negev, Beer-Sheva 84105, Israel}

\author{Malcolm R. Connolly}
\affiliation{Blackett Laboratory, Imperial College London, South Kensington Campus, London SW7 2AZ, United Kingdom}

\author{Eytan Grosfeld}
\affiliation{Department of Physics, Ben-Gurion University of the Negev, Beer-Sheva 84105, Israel}

\author{Eran Ginossar}
\affiliation{Advanced Technology Institute and School of Mathematics and Physics, University of Surrey, Guildford, GU2 7XH, UK}

\begin{abstract}
We introduce a general protocol for obtaining the charge basis density matrix of a superconducting quantum circuit. Inspired by cavity state tomography, our protocol combines Josephson-energy pulse sequences and projective charge-basis readout to access the off-diagonal elements of the density matrix, a scheme we thus dub charge basis tomography. We simulate the reconstruction of the ground state of a target transmon using the Aharonov-Casher effect in a probe qubit to realise projective readout and show the Hilbert-Schmidt distance can detect deviations from the correct model Hamiltonian. Unlocking this ability to validate models using the ground state sets the stage for using transmons to detect interacting and topological phases, particularly in materials where time-domain and spectroscopic probes can be limited by intrinsic noise.  
\end{abstract}
\date{February 2025}

\maketitle
Precise preparation and control of quantum states is key to harnessing the power of entanglement and superposition in quantum-enhanced technologies. 
Quantum state tomography (QST) is a state-of-the-art technique for completely reconstructing quantum states from measurements of an ensemble of identically prepared systems, and has been successfully deployed to study a range of systems including optical photons \cite{Vogel1989,Raymer1994,Leonhardt1995,Leibfired1996}, electron wavepackets emitted by single-electron pumps  \cite{Bisognin2019}, multi-electron atoms with large angular momentum \cite{Klose2001}, and motional states of ions \cite{Poyatos1996}. 
QST is a particularly valuable method for validating and refining physical models used in quantum information processing with highly coherent superconducting cavities \cite{Hofheinz2008,Hofheinz2009,Kirchmair2013}. 
QST of a wider range of superconducting quantum circuits is highly attractive as it could reveal novel dynamics emerging from higher-order harmonics in junctions \cite{Bargerbos2022,willsch_observation_2024} and engineered Josephson potentials in multi-mode circuits \cite{Groszkowski2018}, or detect unique tomographic signatures \cite{Dahan2025} of topological phases \cite{ginossar2014microwave,yavilberg2015fermion,lupo2022implementation,yavilberg2019differentiating}. Existing energy-basis QST protocols, however, are challenging to implement across circuits with variable control parameters and rates of decoherence and relaxation.

In order to overcome this challenge and realize the benefits of tomographic analysis, we propose an inherently more noise-tolerant approach to reconstructing the stationary ground state (GS) of a quantum circuit. 
Evidently such GSs are only non-trivial when reconstructed in an alternative to the energy basis. For a typical target circuit comprising a Cooper pair box with charging energy ($E_C$) and Josephson energy ($E_J$) \cite{Nakamura1999}, when $E_J \ll E_C$ the GS is naturally reconstructed in the relative charge basis (CB). In the most technologically relevant charge-insensitive regime of the transmon ($E_J/E_C\approx100$), however, there are two challenges with projecting the quantum state in the CB. Firstly, unlike the transverse qubit-cavity coupling used for dispersive energy-basis readout, CB readout requires longitudinal coupling to shift the transitions of a probe circuit. Secondly, the CB measurement operator does not commute with the transmon Hamiltonian, generating residual dynamics in the readout signal. 

In this work, we show that fast dynamics can be minimised by bringing the target transmon into the charge regime ($E_J\approx0$), equivalent to quenching gravity in a quantum rotor to reveal the constituent orbitals in the GS \cite{Koch2007}. The reconstruction of the GS density matrix ($\rho_{jk}$) in the CB then proceeds by performing nonlinear least-square regression using a system of equations for $\rho_{jk}$, a well-established technique used for cavity tomography \cite{Hofheinz2008,Hofheinz2009,Kirchmair2013}. To verify the self-consistency and uniqueness of the reconstructed GS we use the Hilbert-Schmitt distance to quantify errors arising from spurious harmonics added to the correct model Hamiltonian. Finally, drawing from recent work \cite{Billangeon2015} we propose longitudinal coupling using the Aharonov-Casher effect \cite{AharonovCasher1984} recently exploited in quantum phase slip devices 
\cite{honigl-decrinisCapacitiveCouplingCoherent2023, degraafDualFraunhoferInterference2020} and flux qubit systems \cite{mooijJosephsonPersistentCurrentQubit1999, orlandoSuperconductingPersistentcurrentQubit1999, Ivanov2001, Friedman2002} where frequencies can be significantly tuned by use of Josephson junctions with reduced tunnelling rate \cite{paauwTuningGapSuperconducting2009, sternFluxQubitsLong2014}. 

The paper is organised as follows. In section \ref{Charge basis tomography}, we present the proposed method: in section \ref{sec:measurement protocol} we focus on the measurement protocol involving the probe qubit and the target system, detailing all the steps, while in section \ref{sec: density matric reconstruction} we derive and explain the density matrix reconstruction protocol, applying it to the case of the transmon model. In section \ref{circuit realisation}, we present a superconducting circuit which can be used to realise the required coupling between the probe and target circuit. The paper concludes with a discussion of possible future use cases for the methods developed here. 

\section{Charge basis tomography}
\label{Charge basis tomography}

In order to reconstruct the density matrix in the CB, we assume that the target system $t$ is formed by a circuit containing a junction under study and characterised by one generalised coordinate only. The probe qubit $p$ is coupled to the target via a term proportional to $\sim\hat{n}\hat{\sigma}_z^{(p)}$, where $\hat{n}$ is the charge degree of freedom of the target circuit and $\hat{\sigma}_z^{(p)}$ is the longitudinal coupling of the probe qubit.
The total Hamiltonian of the coupled target-probe qubit system can be written as
\begin{equation}
\label{Hamj+q}
    H = H_t\left(\hat{n},\hat{\varphi}\right) + g\hat{n}\hat{\sigma}_z^{(p)} + \frac{\Delta_p}{2}\hat{\sigma}_z^{(p)},
\end{equation}
with $H_t\left(\hat{n},\hat{\varphi}\right)$ the Hamiltonian of the target system and $\Delta_p$ the frequency of the probe qubit. If we consider a simple junction model that can be described by only one degree of freedom, its Hamiltonian can be written as, or at least approximated to the form
\begin{equation}
\label{Hamjuncfg}
    H_t\left(\hat{n},\hat{\varphi}\right) = 4E_C(\hat{n}-n_g)^2 + E_Jh_J[\hat{n}_+,\hat{n}_-],
\end{equation}
where $\hat{n}_+ = e^{-i\hat{\varphi}}$ and $\hat{n}_- = e^{+i\hat{\varphi}}$ are the ladder operators associated with the charge operator $\hat{n}$, $n_g$ represents the offset charge parameter of the target system, $h_J$ is a polynomial function of $\hat{n}_\pm$, and $E_C,\,E_J$ are the characteristic charging and Josephson energy associated with the junction. For instance, for the canonical model of the transmon, $h_J[\hat{n}_+,\hat{n}_-] = -(\hat{n}_++\hat{n}_-)/2 = -\cos(\hat{\varphi})$.

\begin{figure}[t]
    \centering
    \includegraphics[width=0.9\linewidth]{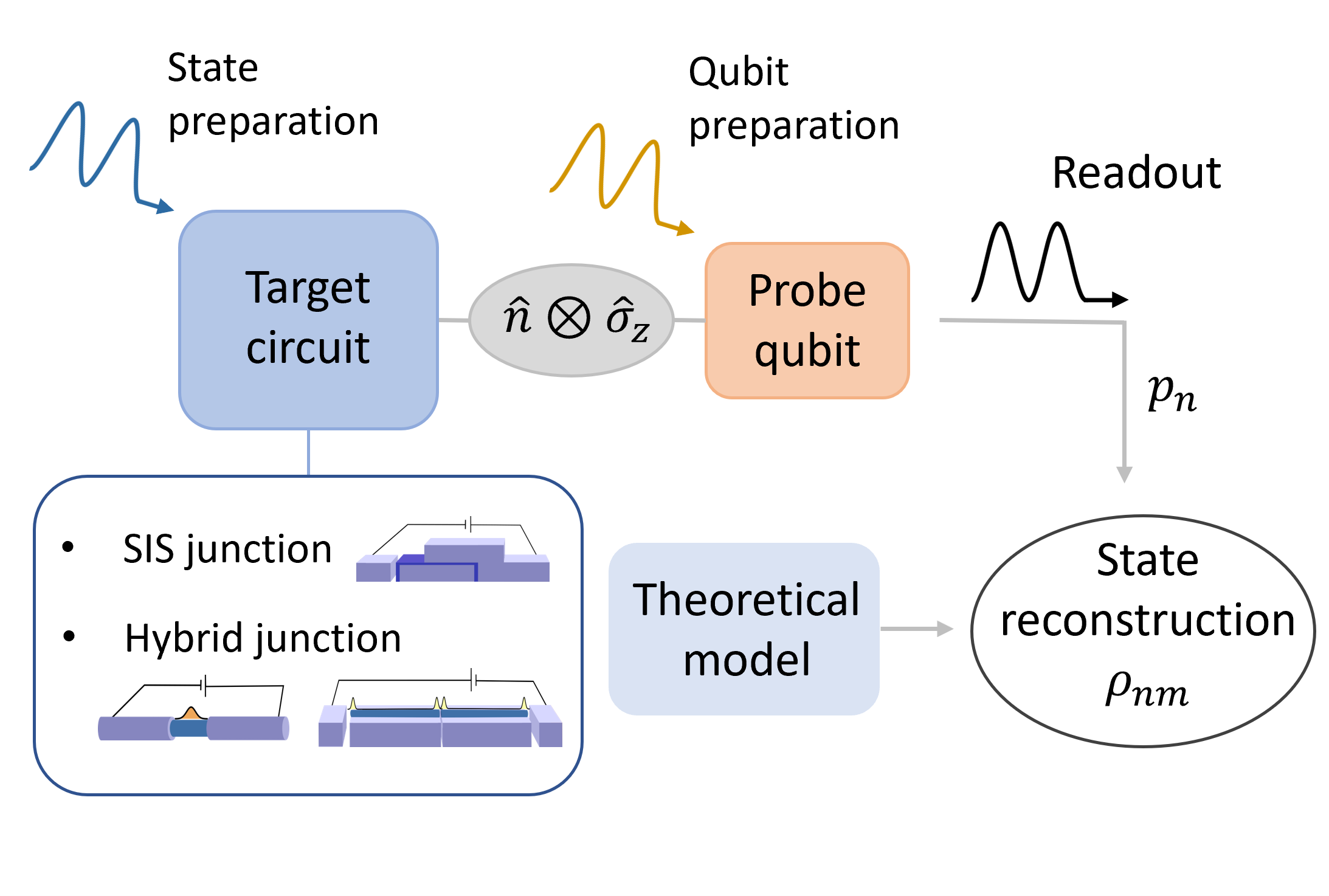}
    \caption{\label{TomDiagram} Schematic showing the protocol for reconstructing the ground state of a superconducting circuit by projective measurements of a coupled probe qubit. The coupling term takes the form $\hat{n}\hat{\sigma}_z$, containing the charge degree of freedom of the target and the longitudinal one of the probe, enables the target ensemble probabilities $p_n=\rho_{nn}$ to be extracted via the projective measurements of the probe qubit. Combining the information obtained from different initial states of the target circuit, a system of equations can be built, connecting the measured diagonal elements of $\rho$ with the off-diagonal ones, and the density matrix of the target system can be reconstructed.}
\end{figure}

Figure \ref{TomDiagram} illustrates our general scheme for reconstructing the density matrix. Known operations applied to the target system lead to controlled mixing (in the sense of unitary evolution) between the components of the density matrix $\rho_{ij}$, leading to a final state $\rho'_{ij}$. The entangled target-probe evolution allows us to use projective measurements performed on the probe qubit to extract the probabilities $p_n=\rho'_{nn}=|\braket{n|\Psi_t}|^2$, with $\ket{\Psi_t}$ the state of the target circuit. These projections are performed in the CB when the coupling is $\sim\hat{n}\hat{\sigma}_z^{(p)}$. This protocol is repeated with different states of the target circuit, so that a system of equations in the density matrix components $\rho_{ij}$ can be constructed and approximatively solved. While in principle the system is infinite, in practice we use a truncated set of basis states. This technique was first applied to trapped atoms in harmonic potentials \cite{Leibfired1996} and more recently to linear and nonlinear superconducting cavities \cite{Hofheinz2009, Kirchmair2013} by conditional measurements of the cavity photon number. Note how an important distinction between our protocol and these previous schemes arises because the relative charge measurement operator $\hat{n}$ does not commute with the Hamiltonian. 
To avoid the fast dynamics resulting from the target-probe interaction, due to the measurement not being QND-like, we implement the protocol described in the following section \ref{sec:measurement protocol}. 
Furthermore, to reconstruct the density matrix the system is adiabatically prepared and measured in its ground state at different values of $E_J=E_J^i$. This is outlined in detail in section \ref{sec: density matric reconstruction}.

\subsection{Measurement protocol}
\label{sec:measurement protocol}
\begin{figure}[t]
    \centering
    \includegraphics[width=0.95\linewidth]{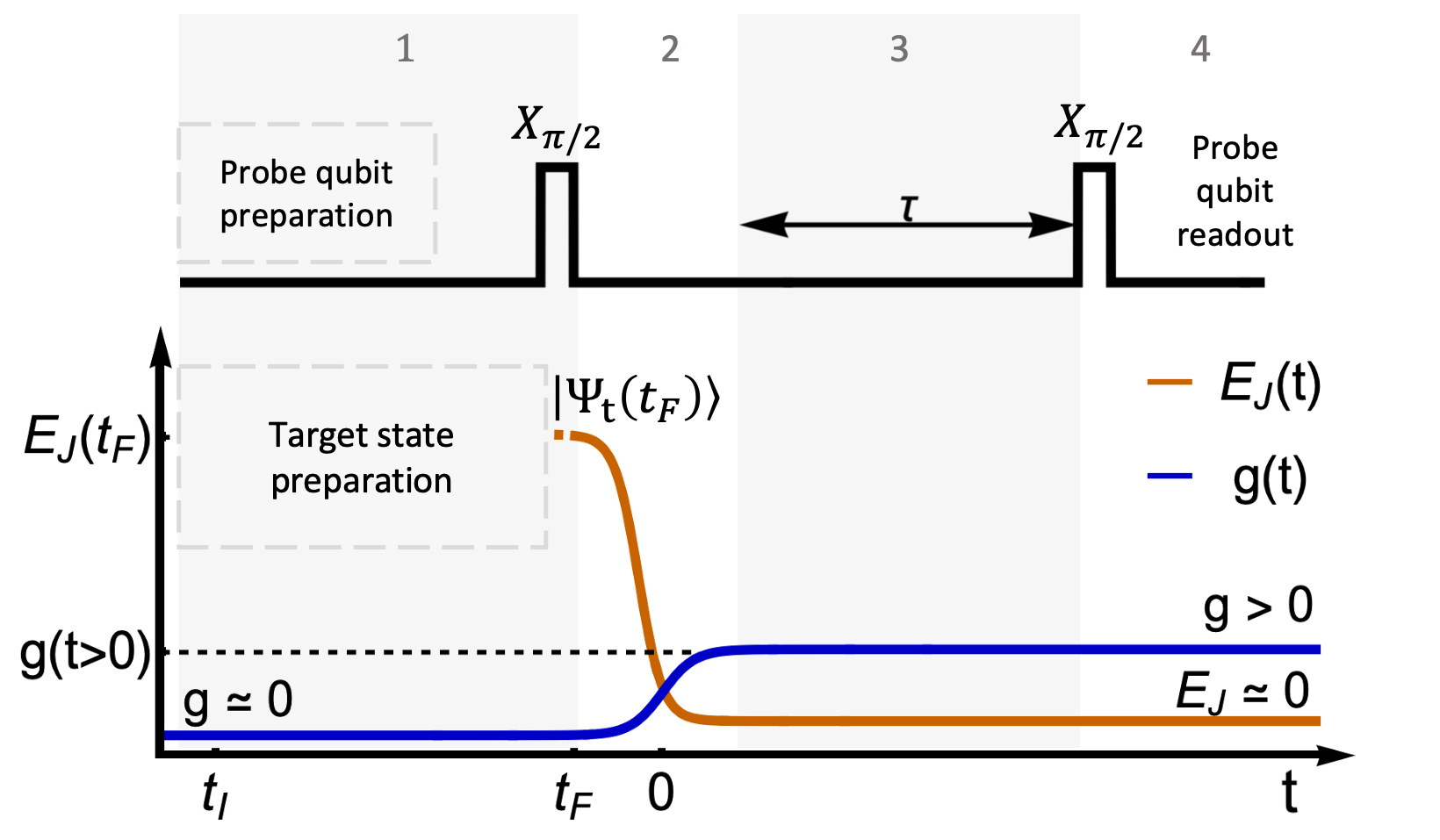}
    \caption{\label{schemeEJg}Schematic of the measurement protocol used to extract the diagonal elements of the target circuit density matrix. For $t_I < t < t_F\lesssim0$, the target and probe are prepared in their initial states (panel no.$1$). In particular, the probe is prepared in an equal superposition of ground and excited states by applying a half $\pi$-pulse. At $t\simeq0$ the coupling is switched on, while the $E_J$ parameter is fluxed down to zero (region 2). This allows the implementation of a Ramsey-like measurement of the probe qubit (regions 3 and 4), whose dynamics is governed by the charge-basis components of the target system.}
\end{figure}

We start by presenting the protocol for extracting the probabilities ($p_n$) of the target system via projective measurements of the probe qubit. In order to obtain the qubit dynamics free from unwanted interactions, the coupling term $g$ is switched on for $t\geq 0$ after the target circuit $\ket{\Psi_t(0)}$ and the probe qubit $\ket{\Psi_p(0)}$ states have been prepared. We combine this with quenching the Josephson energy $E_J$ of the target circuit to switch the system into the charging regime. This can be done experimentally using a symmetrically split Josephson junction, where the effective Josephson energy can be fluxed to zero by an external magnetic field. The protocol is represented in Fig.$\,$\ref{schemeEJg} and can be summarised as follows:
\begin{enumerate}
    \item While $g=0$, for $t_I < t < t_F\lesssim0$ the target circuit is initialised in $\ket{\Psi_t(0)}\simeq \ket{\Psi_t(t_F)}$, at a finite value of $E_J(t_F)>0$. At the same time, using a half $\pi$-pulse the probe qubit is prepared in an equal superposition of its ground and excited states, $\ket{\Psi_p(0)}\simeq \ket{\Psi_p(t_F)}\equiv \frac{1}{\sqrt{2}}\left(\ket{\downarrow} + \ket{\uparrow}\right)$.

    \item At $t\simeq 0$, the Josephson energy $E_J$ of the target circuit is parametrically switched off, while the coupling $g$ is turned on. The speed at which the tuning is implemented needs to be high enough not to affect the system's initial state $\ket{\Psi(0)} =\ket{\Psi_t(0)}\ket{\Psi_p(0)}$.

    \item Similarly to a Ramsey experiment, the probe qubit is left evolving for a time $\tau$ according to Hamiltonian (\ref{Hamj+q}) with $E_J\sim 0$, in which case $H_t$ depends only on the charge operator $\hat{n}$ and the probe qubit dynamics includes the charge components of $\ket{\Psi_t(0)}$, i.e. $c_n = \braket{n|\Psi_t(0)}$.

    \item Another half $\pi$-pulse performed at $t=\tau$ on the probe qubit allows for the projective measurement of the qubit components in the $\sigma_z^{(p)}$-basis.

    \item The repetition of the previous steps at multiple values of $\tau$ leads to the reconstruction of the expectation value of $\hat{\sigma}_x^{(p)}$ as a function of time, $\braket{\hat{\sigma}_x^{(p)}(t)} = \braket{\Psi(t)|\hat{\sigma}_x^{(p)}|\Psi(t)}$,   
    whose Fourier components are proportional to the ensemble probabilities $p_n = |c_n|^2$.
\end{enumerate}
In this scheme, the $E_J$-parameter tuning is required for a proper readout of the charge components. 
When the coupling is switched on, after preparing the combined system in the initial state
\begin{equation}
    \ket{\Psi(0)}\equiv\frac{1}{\sqrt{2}}\ket{\Psi_t}\big(\ket{\downarrow} + \ket{\uparrow}\big),
\end{equation} 
with $\ket{\Psi_t}=\sum_n c_n \ket{n}$, the state $\ket{\Psi(0)}$ evolves according to Hamiltonian (\ref{Hamj+q}). By switching the parameter $E_J$ off while the coupling is being turned on we mitigate the spurious contributions to the qubit dynamics originating from the presence of non-diagonal terms $h_J[\hat{n}_+,\hat{n}_-]$.
Hence we neglect $h_J$ in Eq.$\,$(\ref{Hamjuncfg}) and calculate the evolution of $ \ket{\Psi(t)}$,
\begin{align}
\label{psi_qj}
&{}     \ket{\Psi(t)} = \mathcal{T}e^{-i\hat{H}t}\sum_n \frac{c_n}{\sqrt{2}} \Big(\ket{n, \uparrow} + \ket{n, \downarrow} \Big)
\nonumber\\
&{}     \simeq e^{-iE_C h_C\left(\hat{n}\right)t/\hbar} 
\nonumber\\&{}\hspace{15pt} \cdot
\sum_n \frac{c_n}{\sqrt{2}}  
e^{-i(g\hat{n}\hat{\sigma}_z^{(p)} + \frac{\Delta_p}{2}\hat{\sigma}_z^{(p)})t/\hbar}\Big(\ket{n, \uparrow} + \ket{n, \downarrow} \Big).
\end{align}
From this, the expectation value of the operator $\braket{\hat{\sigma}_x^{(p)}(t)}_\Psi$ takes the form
\begin{align}
\label{sigmaxt}
&{}    \braket{\hat{\sigma}_x^{(p)}(t)}_\Psi = \braket{\Psi(t)|\hat{\sigma}_x^{(p)}|\Psi(t)} \nonumber\\
&{}     =\sum_{m,n} c_m^* \Big(e^{+i(gm + \frac{\Delta_p}{2})t/\hbar}\bra{m, \uparrow} + e^{-i(gm + \frac{\Delta_p}{2})t/\hbar}\bra{m, \downarrow} \Big) \nonumber\\
&{} \quad \cdot c_n\hat{\sigma}_x \Big(e^{-i(gn + \frac{\Delta_p}{2})t/\hbar}\ket{n, \uparrow} + e^{+i(gn + \frac{\Delta_p}{2})t/\hbar}\ket{n, \downarrow} \Big)/2 \nonumber\\
&{}     =\sum_{n} |c_n|^2 \cos\left(\omega_n t\right),
\end{align}
with $\omega_n = \left(\Delta_p + 2gn\right)/\hbar$. When the evolution of $\braket{\hat\sigma_x^{(p)}(t)}_\Psi$ is tracked over time, the probabilities $p_n \equiv |c_n|^2$ can be read in its Fourier transform.

Note that there are several timescales involved in the protocol. First, for the target state's preparation, we have that $t_I\ll 0$ and $t_F\lesssim 0$, with $t_F-t_I$ satisfying the adiabatic condition $t_F-t_I\gg \max_{t_I<t<t_F}\{1/[\epsilon_j(t)-\epsilon_0(t)]\}, \ \forall j\neq0$, with $\epsilon_j(t)$ the eigenstate of $H_t$ at a specific time $t$. We can consider the $E_J$ and $g$ switching done instantaneously about $t=0$. For $0\lesssim t \leq \tau$ the system's state is no longer an eigenstate of Hamiltonian (\ref{Hamj+q}), and is subject to the target system's relaxation and the probe qubit dephasing. Therefore the measurement of $\braket{\hat{\sigma}_x^{(p)}(t)}$ is possible when $\tau < T_1^{(t)},T_\phi^{(p)}$, with $T_1^{(t)}$ and $T_\phi^{(p)}$ the target system's relaxation time and the probe qubit dephasing time, respectively.

\subsection{Density matrix reconstruction}
\label{sec: density matric reconstruction}
In this section we present a procedure for the reconstruction of the state of the target system $H_t$, starting from the measurements on the probe qubit, representing the diagonal elements of its density matrix. In general, to access the off-diagonal elements, controlled $T(\eta)$ operations, with $\eta$ a control parameter, can be engineered and applied to the target system before the probe measurements are performed \cite{Hofheinz2009,Kirchmair2013}. Depending on the transformation, the measured $\rho_{nn}'\equiv p_n$ of the transformed system can be written as a combination of the elements of $\rho_{jk}$ before the transformation,
\begin{align}
\label{rhoformula}
  &{}\rho'_{nn}(\eta) = \braket{n|T(\eta)\rho T(\eta)^\dagger|n} \nonumber\\\
  &{}= \sum_{j,k} \braket{k|T(\eta)^\dagger|n}\braket{n|T(\eta)|j}\rho_{jk} = \sum_{j,k} M_{kj}^n(\eta)\rho_{jk}.
\end{align}
By applying multiple transformations $T(\eta_i)$ and measuring multiple diagonal elements $\rho_{nn}'$, the density matrix of the cavity can be reconstructed, up to a certain truncation $N$. The system of equations obtained using numerous values of $\eta \in \{\eta_i\}$ can be solved if at least one of the applied transformations produces nonzero matrix elements $M_{kj}^n(\eta_i)$ for $|k|,\,|j| <N$. Given that real systems always present some noise, a linear regression model can be used \cite{Kirchmair2013}.

In our protocol, we focus on the reconstruction of the ground state of the target circuit, by probing it at different values of the $E_J$ parameter. The transformation $T(\eta_i=E_J^i)$ can therefore be seen as the time evolution operator obtained by applying an adiabatic pulse $E_J=E_{J,i}(t)$, for $t_I<t < t_F \lesssim 0$, which tunes the Josephson parameter from the value at which we want to reconstruct the density matrix, $E_{J,i}(t_I\ll0) = E_J^0$, to the value at which we measure $\rho_{nn}'$, $ E_J^i = E_{J,i}(t_F)$. The adiabatic transformation is represented by the sum of the projectors connecting the eigenfunctions $\ket{\psi_q^{(t)}[E_J]}$ of the target system $H_t$ with the initial and the final values of $E_J$,
\begin{equation}
    T(E_J^i) \simeq \sum_{q} e^{-i\theta_{i,q}} \ket{\psi_q^{(t)}[E_J^i]}\bra{\psi_q^{(t)}[E_J^0]},
\end{equation}
with $\theta_{i,q} = \int_{t_I}^{t_F} \epsilon_q[E_{J,i}(t')]dt'/\hbar$ the dynamical phase of the eigenstate $q$ accumulated during the adiabatic evolution, and $\epsilon_q[E_J]$ is the eigenenergy, changing parametrically with $E_J$ during the adiabatic tuning. The contribution of these dynamical phases to the tomography process also depends on the prepared target state, as illustrated during the transmon GS reconstruction.
In the adiabatic approximation the matrix $M_{kj}^n \sim M_{kj}^{n,\mathrm{ad}}$ takes the form
\begin{align}
\label{Mmat}
    &{}M_{kj}^{n,\mathrm{ad}}(E_J^i) = \braket{n|T(E_J^i)|j}\braket{k|T(E_J^i)^\dagger|n} \nonumber\\
    &{}\simeq \sum_{p,q}e^{-i[\theta_{i,p}-\theta_{i,q}]}\braket{n|\psi_p^i}\braket{\psi_p^0|j}\braket{k|\psi_q^0}\braket{\psi_q^i|n},
\end{align}
where we have introduced the shortened notation $\ket{\psi_q^0}\equiv \ket{\psi_q^{(t)}[E_J^0]}$ and $\ket{\psi_q^i}\equiv \ket{\psi_q^{(t)}[E_J^i]}$.\\

\begin{figure}[t!]
      \centering
\includegraphics[width=\linewidth]{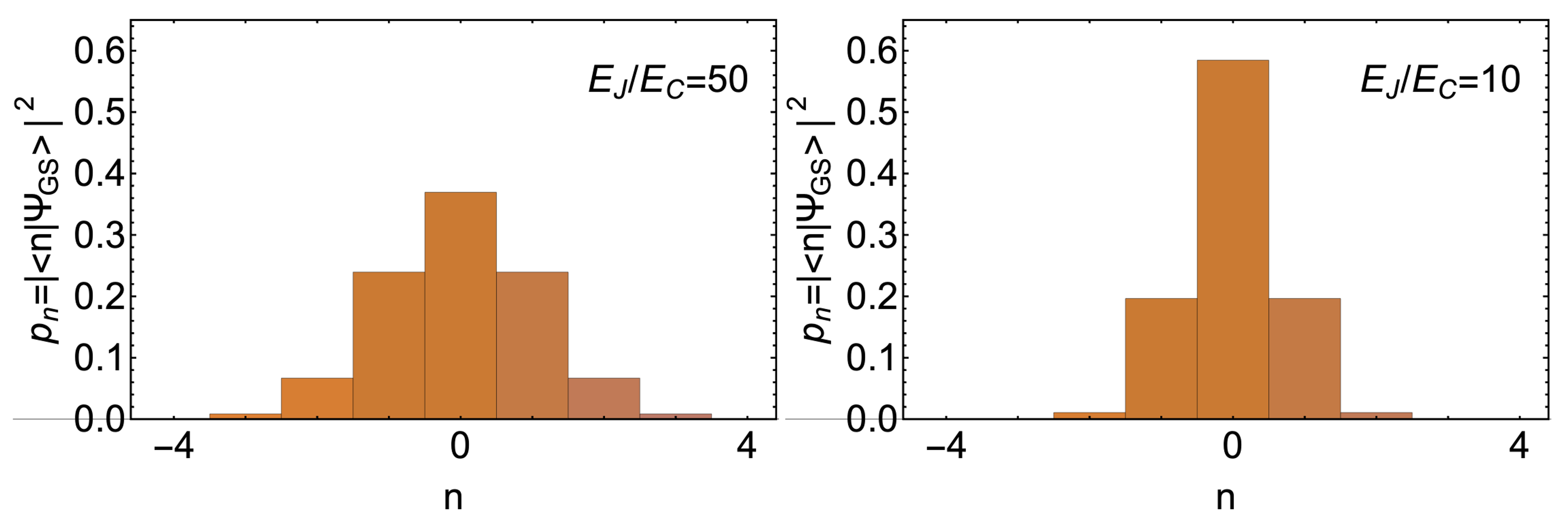}
\caption{\label{GSProps}Charge probability components $p_n = |\braket{n|\psi_{0}^{(T)}}|^2$ of the ground state of the transmon, at $E_J/E_C=50$ (left) and $E_J/E_C=10$ (right).} 

\end{figure}

\noindent\textbf{Transmon case.} As an example, we determine the system of equations obtained when probing a simple system consisting of a Josephson junction-based superconducting island (charge-based qubit) in the regime of high Josephson energy $E_J \gg E_C$ (transmon regime). In this case, the target Hamiltonian can be written as
\begin{align}
\label{Htransmon}
    H_t&{}\equiv H_T = 4E_C(\hat{n}-n_g)^2 -E_J\cos(\hat{\varphi}),
\end{align}
where $n_g$ is the charge offset of the superconducting island written in terms of the number of Cooper pairs. $E_J$ can be representing an effective Josephson energy of a split junction, thus it can depend on an external flux parameter, i.e. $E_J = E_J(2\pi f_t)$. The solutions of Eq.$\,$(\ref{Htransmon}) are superpositions of Mathieu's functions \cite{Cottet2002}, and in the regime where $E_J\gg E_C$ they can be approximated by the solution of a quantum harmonic oscillator, with frequency $\omega_T \simeq \sqrt{8E_CE_J} -(E_C/2)$ \cite{Koch2007},
\begin{align}
    \psi_k^{(T)}(\varphi) &{}\simeq \psi_k^{h.o.}(\varphi) = \braket{\varphi|\psi_k^{h.o.}} \nonumber\\
    &{}= \mathcal{N}_k(E_C,E_J) e^{-in_g\varphi}e^{-(\beta\varphi)^2/2}\mathrm{He}_k(\beta\varphi),
\end{align}
where $ \mathcal{N}_k(E_C,E_J) = (1/\sqrt{2^k k!})(E_J/8\pi^2 E_C)^{1/8}$, $\beta = (E_J/8E_C)^{1/4}$ and $\mathrm{He}_k(x)$ is the $k$th-order Hermite polynomial. For the ground state it just corresponds to a Gaussian function with $\sigma = 1/\beta$. Therefore, to obtain a good resolution of the ground state, its expansion in the charge basis $\ket{\psi_{GS}^{(T)}}=\ket{\psi_{0}^{(T)}}=\sum_n c_n^0\ket{n}$ has to contain at least all the terms with $-N\leq n\leq N$, with truncation $N\sim 3\beta = 1.77(E_J/E_C)^{1/4}$. For a circuit with $E_J/E_C \sim 50$, this corresponds to $N\sim5$ (See Fig.$\,$\ref{GSProps}).
\\
The eigenstates' coefficients in the charge basis, appearing in Eq.$\,$(\ref{Mmat}), can be written as
\begin{align}
\label{cnEigCoef}
    c_n^k =&{} \braket{n|\psi_k^{(T)}} = \frac{1}{2\pi}\int_{-\pi}^{\pi} \braket{n|\varphi}\braket{\varphi|\psi_k^{(T)}}d\varphi \nonumber\\
    \simeq&{}\ \frac{\mathcal{N}_k}{2\pi}\int_{-\infty}^{+\infty} e^{i(n-n_g)\varphi}e^{-(\beta \varphi)^2/2}\mathrm{He}_k(\beta\varphi)d\varphi,
\end{align}
Equation (\ref{cnEigCoef}) has the form of a Fourier transform of a function of $z=\beta\varphi$, computed at $(n-n_g)/\beta$. For example, for $k=0$ (the GS), $c_n^0$ is a Gaussian function,
\begin{align}
\label{cnEigCoefGS}
    c_n^0 \simeq&{} \frac{\mathcal{N}_0}{2\pi}\int_{-\infty}^{+\infty} e^{i(n-n_g)\varphi}e^{-(\beta \varphi)^2/2}d\varphi\nonumber\\
    =&{}\ \frac{\mathcal{N}_0}{\beta\sqrt{2\pi}} e^{-(n-n_g)^2/2\beta^2}.
\end{align}
For a generic $k$, the coefficients $c_n^k$ can be determined by taking the Fourier transform of the completeness relation for the Hermite polynomials,
\begin{equation}
\label{completeEqHermite}
    e^{-\frac{x^2}{2}+2xt-t^2} = \sum_{k=0}^{+\infty} e^{-\frac{x^2}{2}} \mathrm{He}_k(x) \frac{t^k}{k!}.
\end{equation}
Let's start by integrating both sides of the equation:
\begin{equation}
\label{completeEqHermite_integrated}
    \int e^{i\omega x} e^{-\frac{x^2}{2}+2xt-t^2} dx = \sum_{k=0}^{+\infty}\int e^{i\omega x} e^{-\frac{x^2}{2}} \mathrm{He}_k(x) \frac{t^k}{k!} dx.
\end{equation}
The terms on the right-hand side of the equation above are the expressions we need, while the left-hand side takes the form
\begin{align}
\label{completeEqHermite_integrated2}
    &{}\int e^{i\omega x}e^{-\frac{x^2}{2}+2xt-t^2}dx = \int e^{-\frac{1}{2}\left[ x^2 -2x(i\omega+2t) +2t^2\right]}dx \nonumber\\
    &{}=e^{\frac{1}{2}(i\omega+2t)^2-t^2}\int e^{-\frac{1}{2}\left( x - 2t -i\omega\right)^2}dx \nonumber\\
    &{}= \sqrt{2\pi}\,e^{-\omega^2/2 + 2i\omega t+t^2} \nonumber\\
    &{}= \sqrt{2\pi}\sum_{k=0}^{+\infty} e^{-\frac{\omega^2}{2}} \mathrm{He}_k(\omega) \frac{(it)^k}{k!}.
\end{align}
Equating the coefficients of the two sums found in equations (\ref{completeEqHermite_integrated})-(\ref{completeEqHermite_integrated2}), and substituting $\omega=(n-n_g)/\beta$, the expression for $c_n^k$ is obtained:
\begin{align}
    c_n^k =\frac{i^k}{\sqrt{2^{k+1} k!\pi^{3/2}\beta\,}} e^{-\frac{(n-n_g)^2}{2\beta^2}} \mathrm{He}_k\left[\frac{n-n_g}{\beta}\right],
\end{align}
where we have replaced $\mathcal{N}_k = \sqrt{\beta/2^k k!\sqrt{\pi}}$. 
Using this expression, we can explicitly write the matrix elements $M_{kj}^{n,\mathrm{ad}}$ defined in equation (\ref{Mmat}), as
\begin{figure*}[t!]
      \centering
    \includegraphics[width=0.9\linewidth]{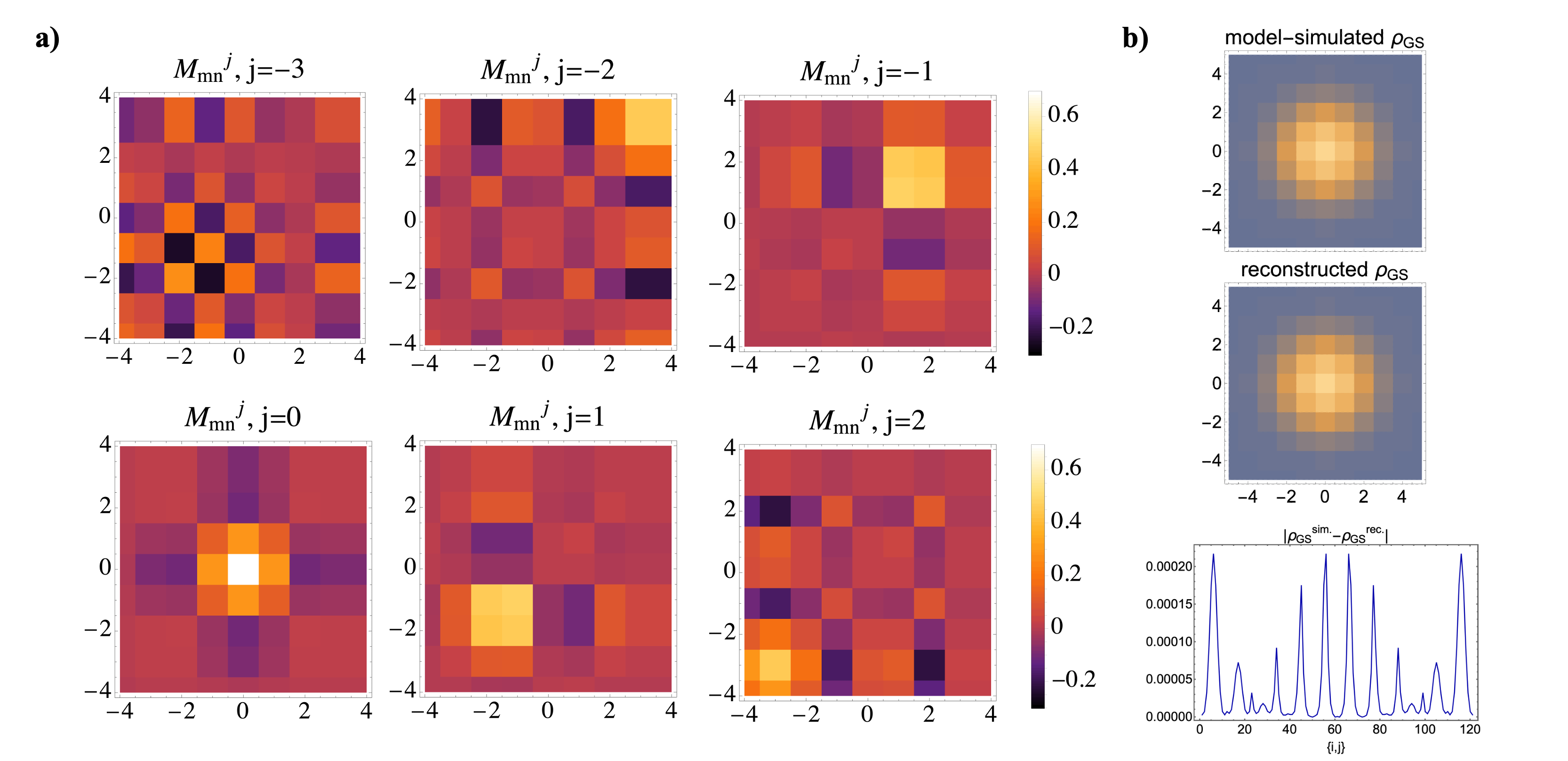}
    \caption{\label{MatrixElementsPlusLinRegGS}Example of application of the tomography protocol to reconstruct the ground state density matrix of the transmon. a) Matrix elements $M_{nm}^j[E_J^i]$, with $E_J^i=10E_C$, used for the reconstruction of the ground state's density matrix at $E_J/E_C=50$; b) Comparison between the ground state determined numerically from the Hamiltonian and the reconstruction using the least-square regression method and $21$ configuration points with $10\leq E_J^i/E_C \leq 50$. The bottom panel shows the absolute value of the difference between each element of the two density matrices.} 
\end{figure*}
\begin{widetext}
\begin{align}
    \label{MmatTransmon}
    &{}M_{kj}^{n,\mathrm{ad}} \simeq \sum_{p,q}e^{-i[\theta_{i,p}-\theta_{i,q}]}
    \, c_n^p[E_J^i] \, c_j^{p}[E_J^0]^* \, c_k^q[E_J^0] \, c_n^{q}[E_J^i]^* \nonumber\\
    &{}\simeq 
    \frac{e^{-\frac{(n-n_g)^2}{\beta_i^2}}
    e^{-\frac{(j-n_g)^2}{2\beta_0^2}}
    e^{-\frac{(k-n_g)^2}{2\beta_0^2}} }{4\pi^3\beta_0\beta_i} 
    \left\{ \sum_{p} 
    \frac{e^{-ip\theta_{i,0}}}{2^{p} p!}  \,\mathrm{He}_p\left[\frac{n-n_g}{\beta_i}\right]
    \mathrm{He}_p\left[\frac{j-n_g}{\beta_0}\right] \right\}
    \left\{ \sum_{q} 
    \frac{e^{iq\theta_{i,0}}}{2^{q} q!}  \,\mathrm{He}_q\left[\frac{k-n_g}{\beta_0}\right]\mathrm{He}_q\left[\frac{n-n_g}{\beta_i}\right] \right\},
\end{align}
\end{widetext}
with $\theta_{i,0} = \int_{t_I}^{t_F}\omega_T[E_{J,i}(t')]dt'$.
From the formula above it is clear that the adiabatic transformation mixes the components of the wavefunction in a non-trivial way. Fig.$\,$\ref{MatrixElementsPlusLinRegGS}a shows, for instance, the matrix elements associated with the case $E_J^0/E_C=50$, $E_J^i/E_C=10$ and $n_g=0$. Because of the common scaling factors at the beginning of Eq.$\,$(\ref{MmatTransmon}), the matrix elements are nonzero only for $|n|\leq 3\beta_i/\sqrt{2} \sim 2$ and $|k|,|j|\leq 3\beta_0 \sim 5$. Note that the resolution of the probabilities is reduced because we can only measure the probabilities of the wavefunction. \vspace{2pt}\\
\\
\noindent\textbf{Least square regression.} We want to find a solution, or at least the best approximated solution, to the system of equations Eq.$\,$(\ref{rhoformula}), $\rho'_{nn}(E_{J,i}) = \sum_{j,k} M_{kj}^n(E_J^i)\rho_{jk}$, with $\rho_{jk}$ the unknown variables and $\{E_J^i\}$ a set of system configurations with different values of $E_{J}=E_J^i$.
We again focus on the reconstruction of the GS of the system, assuming that we only need a finite truncation for its representation in the charge basis, $\mathrm{dim}(\rho) = (2N_0+1)\times(2N_0+1) \sim 12\sqrt{E_J^0/E_C}$. At the same time, the measured probabilities $\rho'_{nn}(E_J^i)$ are only visible for $|n|\leq N_i \sim (E_J^i/E_C)^{1/4}$. Assuming $E_J^i < E_J^0$, to be able to solve the system of equations we need to measure the system at several configurations $i=1,\dots i_{\mathrm{max}}$, with 
\begin{equation}
    i_{\mathrm{max}} \gtrsim\, \mathrm{dim}(\rho)/2\mathrm{max}|n| \sim 12\left( \frac{(E_J^0)^2}{E_J^iE_C} \right)^{\frac{1}{4}}.
\end{equation}
In this way the system is overdetermined and an approximate solution can be found using a least squares regression model. To ensure that the physical constraints of the density matrix are satisfied, the hermitianity of $\rho$ and its normalisation $\mathrm{Tr}{\rho}=1$ can be imposed by parametrising its elements $\rho_{jk}$, while its positive-semidefinite property can be fulfilled by truncating its negative eigenvalues after the simulation.

As an example, we reconstruct the transmon GS density matrix in the case of $E_J/E_C=50$. This state can be reconstructed with a truncation of $N_0\sim5$, thus $\rho$ has a dimension of $11\times11$. When we probe a state at, for instance,  $E_J^i/E_C\sim10$ the resolution allows us to measure $\rho'_{nn}(10E_C)$ up to $n = N_i\sim 2$, giving us $5$ measurements of the diagonal elements. We need at least about $i_{\mathrm{max}}\sim 25$ configurations $\{E_J^i:\ 10 \leq E_J^i/E_C \leq 50\}$ to be able to reconstruct the density matrix $\rho[E_J^0]$. This number may be lower depending on how many probability values can be measured for each configuration. We run a simulation of measurements, with zero-noise error, of $\rho'_{nn}[E_J^i]$ at $21$ configuration points, with $10 \leq E_J^i/E_C \leq 50$, and perform a least-squares regression using Eqs.$\,$(\ref{rhoformula}) and (\ref{Mmat}). For the latter, since we are probing an eigenstate of the system the accumulated dynamical phases $\theta_{i,p}=\int_0^t \epsilon_p[E_{J,i}(\tau)]d\tau$ do not contribute to the final density matrix, thus we set them to zero. For the reconstruction of an arbitrary superposition of eigenstates, the time spent during the adiabatic transformation must be taken into account, and the dynamical phases cannot be omitted. It can be seen from the lower panel of Figure \ref{MatrixElementsPlusLinRegGS}b that the least-square regression model is able to reconstruct the ground state of the system within a fifth-digit error.

\begin{figure}[b]
      \centering
    \includegraphics[width=0.9\linewidth]{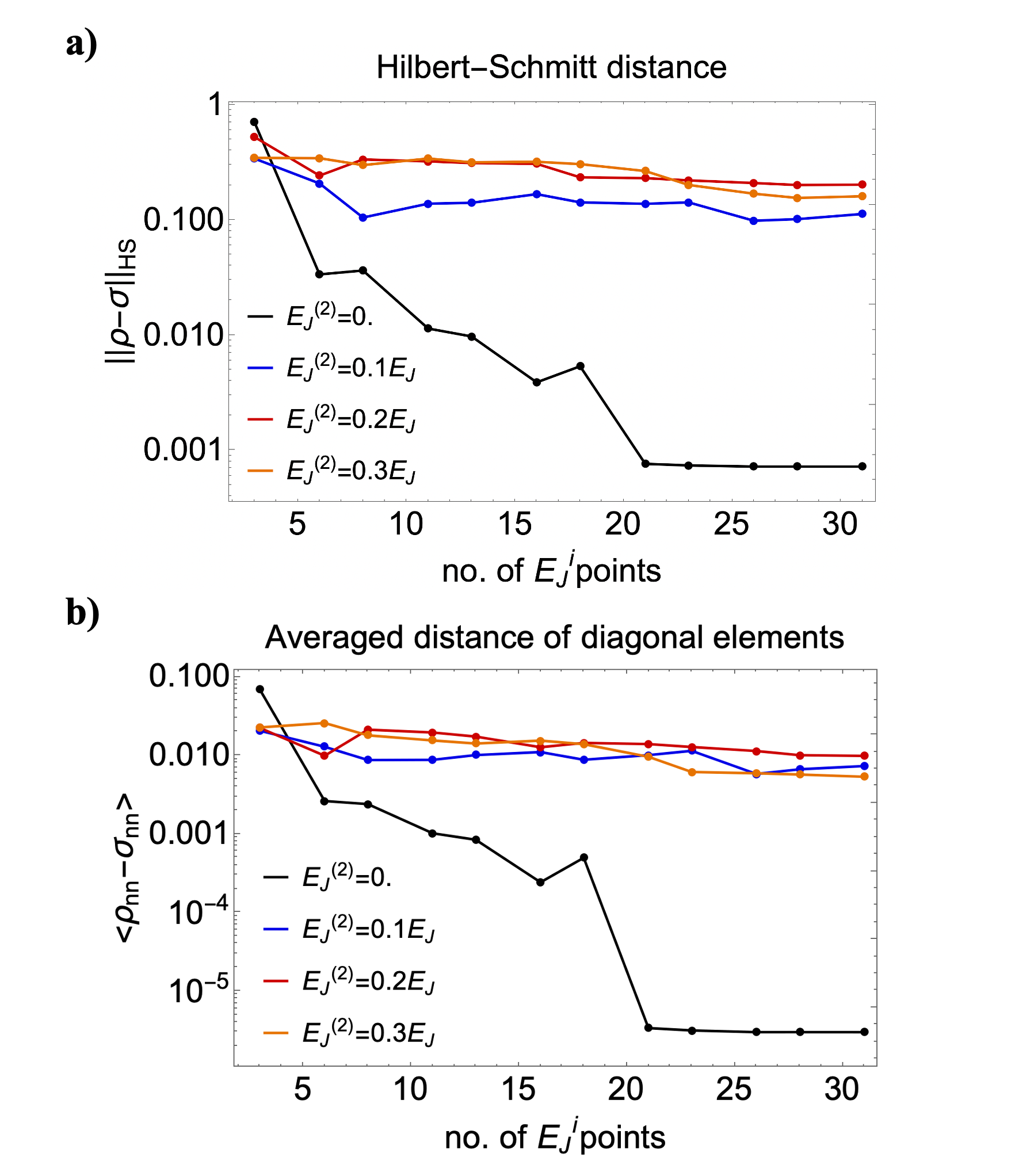}
    \caption{\label{HSdistance}Reconstruction of the GS density matrix $\rho$ and comparison with the expected GS density matrix $\sigma$ from the assumed model, using (a) the Hilbert-Schmitt distance, $||\rho-\sigma||_{HS}= \sum_{ij}|\rho_{ij}-\sigma_{ij}|^2$, and (b) the average distance between their diagonal elements, $\braket{\rho-\sigma}_{nn} = \sum_n(\rho_{nn}-\sigma_{nn})/\mathrm{dim}(\rho)$, vs the number of sampled $E_J^i$ points. The ``measured" diagonal elements are simulated with the same model Hamiltonian of a transmon with a $\cos(\varphi)$ potential, while the reconstruction is done with a model that includes an additional $E_J^{(2)}\cos(2\varphi)$ potential. As the number of sampled $E_J^i$ points increases, the mismatch between the model used and the actual Hamiltonian of the system becomes more apparent, not bringing improvement in the density matrix reconstruction.} 
\end{figure}

In addition, we want to test the self-consistency of the method by reconstructing the transmon GS density matrix, Eq.$\,$(\ref{Htransmon}), using a different model $H_s$ for the calculation of the $M_{kj}^{n,ad}(E_J^i)$ matrix elements, Eq.$\,$(\ref{Mmat}). Fig. \ref{HSdistance} shows the results of this simulation, taking as comparison the measure of the Hilbert-Schmitt distance, $||\rho^{rec}-\rho^{sim}|| = \sum_{ij}|\rho_{ij}^{rec}-\rho_{ij}^{sim}|^2$, and the average distance between the diagonal elements, $\braket{\rho^{rec}_{jj}-\rho^{sim}_{jj}} = \sum_{j}(\rho_{jj}^{rec}-\rho_{jj}^{sim})/\mathrm{dim}(\rho^{rec})$, between the reconstructed $\rho^{rec}$, and the $\rho^{sim}$ expected from the simulation, i.e. calculated with $H_s$. The alternative model is represented by the transmon Hamiltonian with an added $\cos(2\varphi)$ nonlinearity, 
\begin{equation}
    \label{HmodelCos2Phi}
    H_s = 4E_C(\hat{n}-n_g)^2 -E_J\cos(\hat{\varphi}) -E_J^{(2)}\cos(2\hat{\varphi}),
\end{equation}
that can represent additional harmonic structure in the energy-phase relation within the junction. We test how the model depends on variations of $E_J^{(2)}/E_J$ and the number of sampled points. As can be seen from the plots in Figs. \ref{HSdistance}(a)-\ref{HSdistance}(b), increasing the number of sampled points used for the fit clearly discriminates between the model used to represent the actual GS and the models with the added $\cos(2\varphi)$ nonlinearity, suggesting a potential procedure for testing different models.
\\
\\
\noindent\textbf{Residual $E_J$ contribution to the $p_n$ measurement.} 
\begin{figure}[t]
      \centering
    \includegraphics[width=0.95\linewidth]{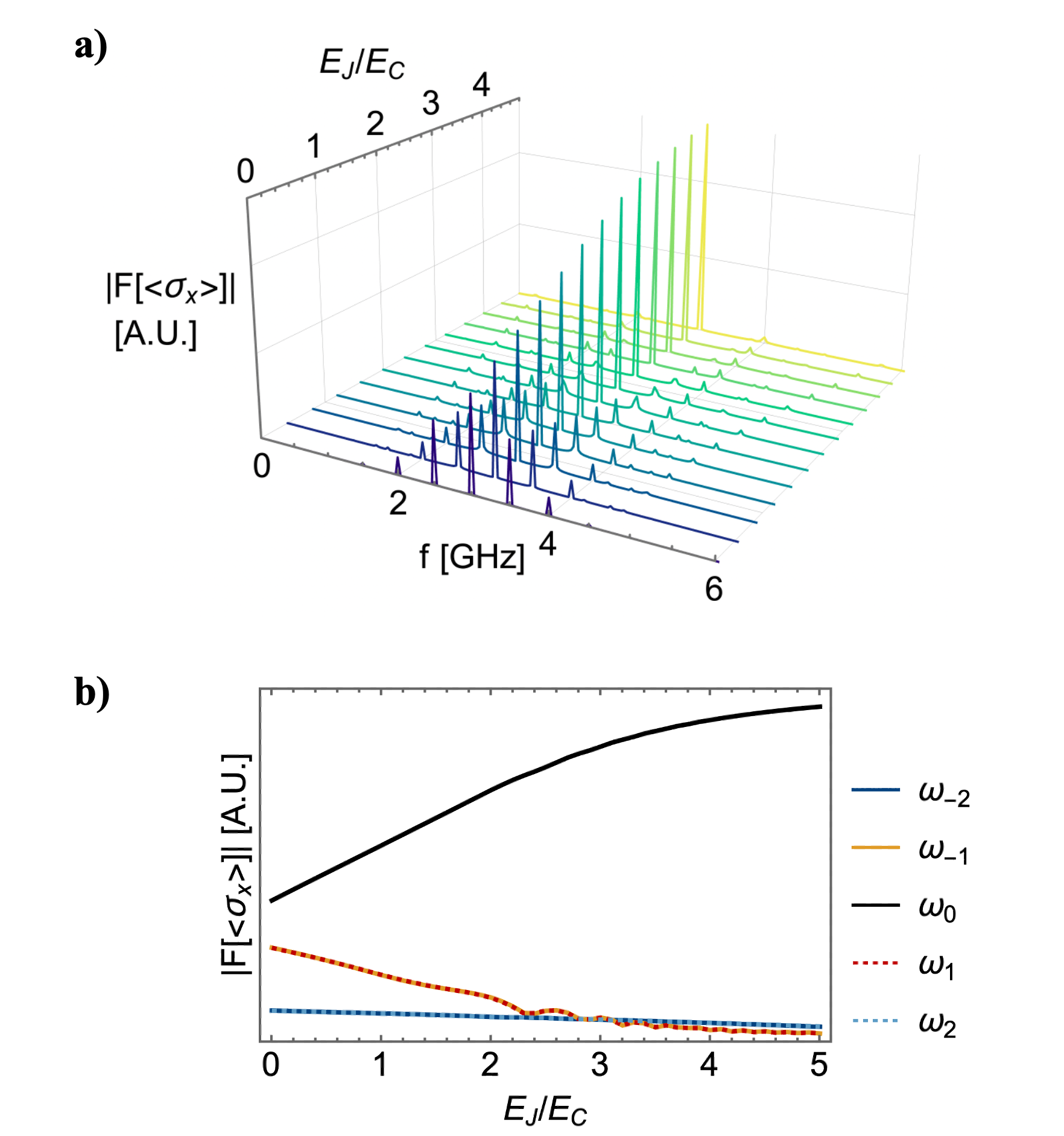}
    \caption{\label{finiteEJpeaks}Effect of residual $E_J$ contributions to $\braket{\hat\sigma_x^{(p)}}$ measurement in the case where the target circuit is a transmon: (a) Spectral profile of $\braket{\hat\sigma_x^{(p)}}$ for different $E_J/E_C$ values; (b) Peaks heights decrease with varying $E_J/E_C $. The central peak is found at $\omega_0/2\pi = \Delta_p/h = 3\ \mathrm{GHz}$, while the other ones are found at $\omega_n/2\pi = (\Delta_p + 2gn)/h$ at $E_J=0$, but they start to deviate from these values for $E_J/E_C\gg 1$. In this range, the peaks heights also experience a decrease in magnitude which can affect their resolution during the measurement. } 
\end{figure}
The least-square regression used to reconstruct the transmon's GS relies on the knowledge of the diagonal terms $\rho_{nn}$ that are extracted from the measurement protocol outlined in Sec \ref{sec:measurement protocol}. The latter requires switching off the Josephson energy parameter, $E_J$, which in realistic setups can only be switched to a nonzero value within the charging regime, $E_J \ll E_C$. A residual value of $E_J$ can partially spoil the resolution of the Fourier transform peaks, reducing their magnitude or introducing new ones. We have analysed this scenario by numerically simulating the dynamics of $\braket{\hat\sigma_x^{(p)}}$, whose frequency components reflect the contribution of each charge state to the target GS.
In Fig.$\,$\ref{finiteEJpeaks}, we simulate $F[\braket{\hat\sigma_x^{(p)}}]$ when a nonzero value of $E_J$ is present during the measurement process ($t\geq 0$), and plot it together with the magnitude of the peaks as this residual $E_J$ is varied. We observe that peak frequencies and magnitudes are unaffected by small corrections for $E_J/E_C \ll 1$, and argue that the methods introduced here are valid even when a small residual $E_J$ is included in the numerics.

\section{Circuit realisation}
\label{circuit realisation}
\begin{figure}[t]
    \centering
    \includegraphics[width=\linewidth]{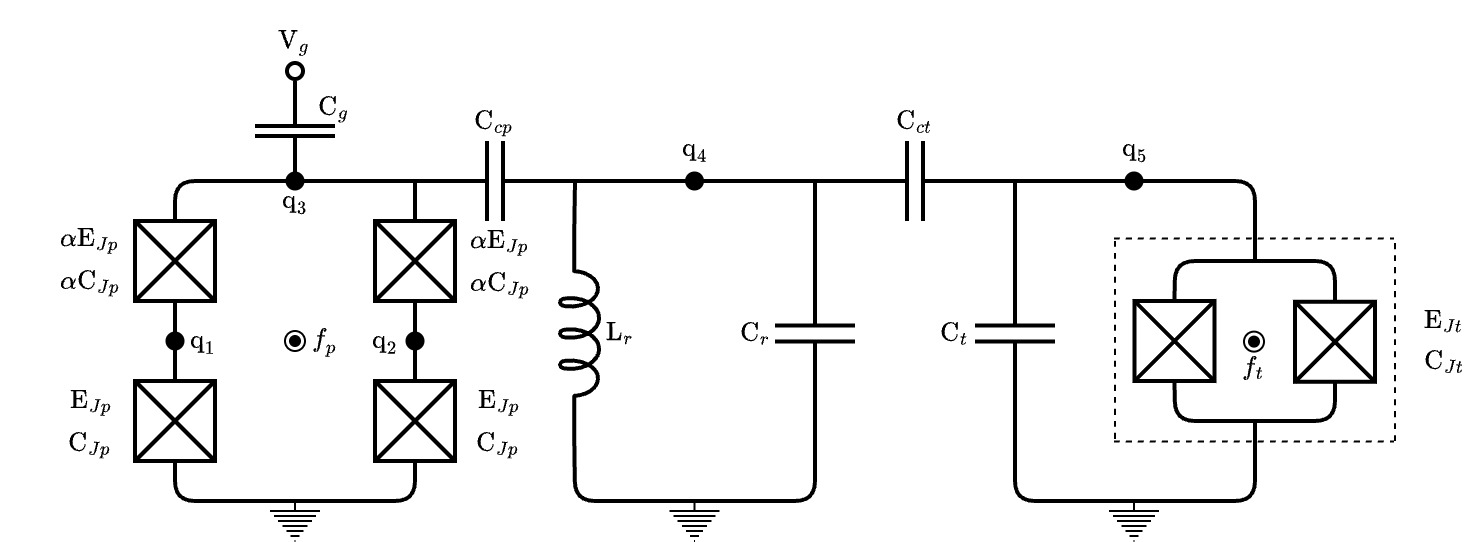}
    \caption{\label{resonator_coupled_circuit} Diagram of the resonator-coupled circuit. The loop on the left is a four-junction flux qubit, here referred to as the probe qubit. The upper two junctions are scaled by a factor of $\alpha$ relative to the size of the lower two junctions. Flux is threaded through the circuit, denoted by $f_p$, allowing the circuit to be biased to the half-flux point. In addition there is a gate charge applied to the node $q_3$, resulting in a charge offset defined as $n_g^{(p)}$. The probe qubit is capacitively coupled to the resonator, shown as the middle circuit. The resonator consists of an inductor $L_r$ and a capacitor $C_r$, connected in series. Finally the resonator is also capacitively coupled to the target qubit, the loop on the right. The target qubit is a tunable Cooper pair box, consisting of a capacitor in series with two Josephson junctions connected in parallel. Within the coupling simulations the tunable junction was treated as a single junction for simplicity.}
\end{figure}

\begin{figure}[b]
    \includegraphics[width=\linewidth]{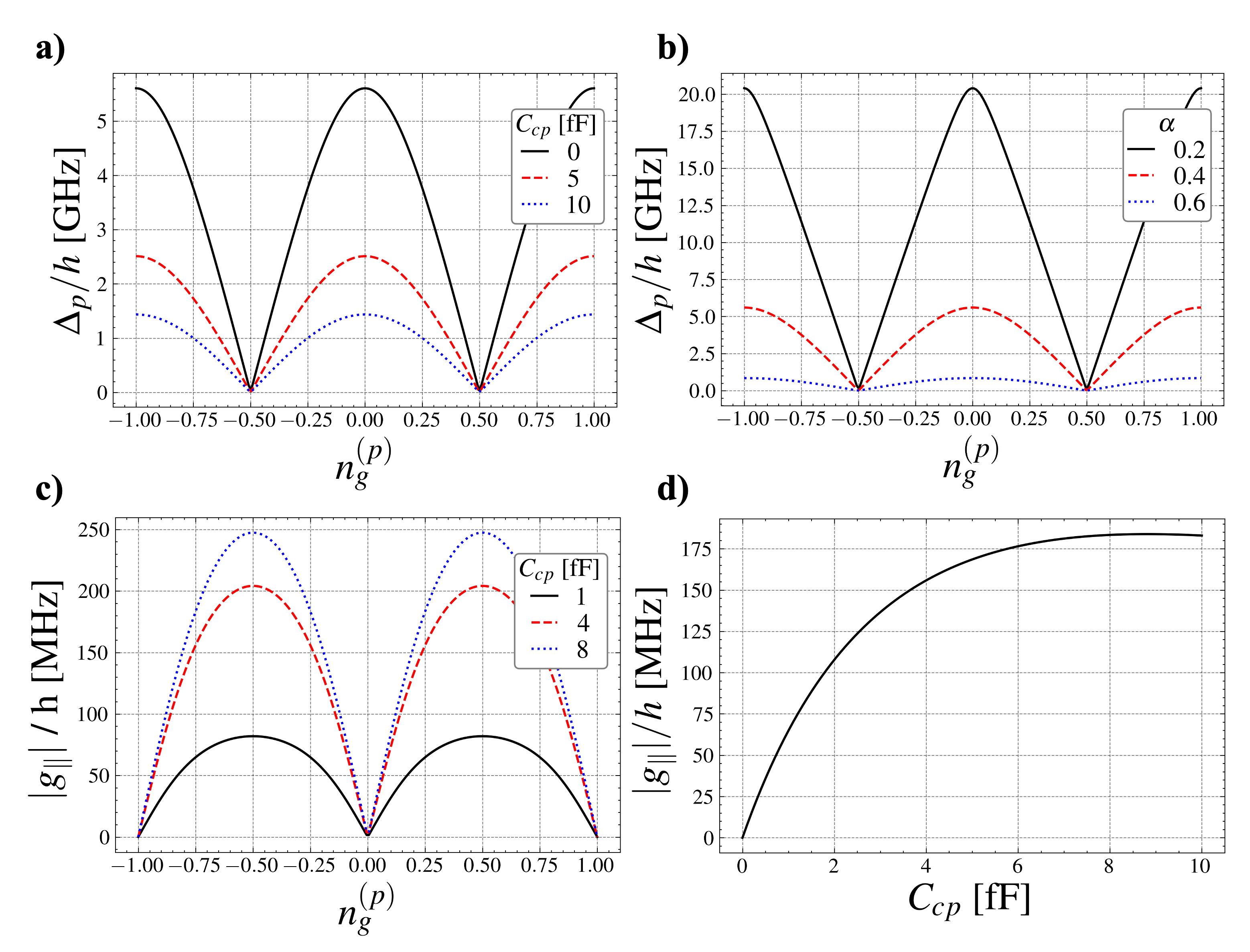}
    \caption{Probe qubit energy gap and longitudinal coupling to the resonator in the circuit described by Fig.$\,$\ref{resonator_coupled_circuit}: (b) Probe qubit energy gap as a function of the charge offset parameter $n_g^{(p)} = q_3/(2e)$, for a range of values of the probe-resonator coupling capacitance $C_{cp}$, with $\alpha=0.4$; (b) Probe qubit energy gap as a function of $n_g^{(p)}$, for a range of values of the asymmetry parameter $\alpha$, with $C_{cp}=0$; (c) Probe qubit-resonator longitudinal coupling strength as a function of $n_g^{(p)}$, for a range of values of the probe-resonator coupling capacitance $C_{cp}$, with $\alpha=0.4$; (d) Probe qubit-resonator longitudinal coupling strength as a function of the coupling capacitance $C_{cp}$, with $n_g^{(p)} = 0.25$ and $\alpha=0.4$. The other circuit parameters are set to: $f_p=0.5$,  $E_{Jp}/h=121$ GHz, $E_{Jt}/h=5$ GHz, $C_{Jp}=8$ fF, $C_{Jt}=4$ fF, $C_t=40$ fF, $C_{ct}=5$ fF, $C_r=100$ fF, $L_r=10$ nH.}
    \label{resonator_coupled_circuit_parameter_graphs}
\end{figure}

We now propose a possible circuit realisation of the required longitudinal coupling $\sim\hat{n}\hat{\sigma}_z$ using a $\hat{\sigma}_z$ operator acting in the subspace of a four-junction flux qubit. Previous works on two- and three-junction SQUID rings \cite{Ivanov2001, Friedman2002} have shown that the control of the accumulated charge in the superconducting island formed by the junctions can be used to engineer the spectrum of a flux qubit. This phenomenon is related to the Aharonov-Casher (AC) effect \cite{AharonovCasher1984} and can also be exploited in circuits where the accumulated charge in the island represents an additional degree of freedom. For instance, when a capacitive coupling is added via one of the superconducting islands. Since the flux through a split junction with reduced Josephson energy can be used to tune the flux qubit frequency, a capacitive coupling connected to this split junction can be used to implement the required interaction term \cite{Billangeon2015}. Fig.$\,$\ref{resonator_coupled_circuit} shows a schematic of the proposed circuit. It consists of a loop of four junctions, representing our probe qubit, capacitively coupled via a resonator to the target circuit, represented in this case by a transmon qubit. The flux qubit has two junctions of reduced Josephson energy $\alpha E_{Jp}$, $\alpha<1$, controlling the qubit frequency. The coupling capacitance is connected to the superconducting island created by these two $\alpha$-junctions, while an external gate voltage applied to the same island is used to tune the coupling and the qubit frequency by exploiting the AC interference. 
Moreover, the use of a resonator to mediate the coupling enables the probe-target coupling strength to be switched, which is required for the tomography protocol. 
When there is no asymmetry between the $\alpha$-junctions and at the flux qubit's sweet-spot $f_p = 1/2$, $\Phi_p^\mathrm{ext}=2\pi f_p$ being the external magnetic flux threading through the qubit loop, any transverse coupling between the probe qubit and the resonator vanishes. In the qubit computational subspace, the Hamiltonian can then be written as (see also Appendix \ref{Appendix_circuit_analysis} for a detailed derivation):
\begin{align} \label{eq:resonator_coupled_hamiltonian_eb0}
        \hat{H} = &\, 4E_{Ct}(\hat{n}^{(t)})^2+E_{Jt}[1-\cos(\hat{\varphi}^{(t)})] + \frac{\Delta_p}{2} \hat{\sigma}^{(p)}_z + \hbar\omega_r  \hat{a}^\dagger \hat{a} \nonumber\\
        - & i g^{(pc)}_{\parallel} \hat{\sigma}^{(p)}_z (\hat{a}^\dagger - \hat{a}) + g^{(pt)}_{\parallel} \hat{\sigma}^{(p)}_z \hat{n}^{(t)} - i g^{(ct)}_{\perp} (\hat{a}^\dagger - \hat{a}) \hat{n}^{(t)},
\end{align}
where $\hat{n}^{(t)}=q_5/2e$ and $\hat{\varphi}^{(t)}=2\pi\phi_5$ are the transmon charge and flux operators, $\hat{\sigma}^{(p)}_k$ are the Pauli matrices associated with the flux qubit operational subspace, $\Delta_p$ is the flux qubit energy gap, $\hat{a}^\dagger$ and $\hat{a}$ are the resonator bosonic operators, $\omega_r$ is the resonator frequency; $g^{(pc)}_{\parallel}$ and $g^{(pt)}_{\parallel}$ are the amplitudes of the resulting flux qubit-resonator and flux qubit-transmon longitudinal couplings, respectively; finally, $g^{(ct)}_{ \perp}$ is the transverse coupling between the resonator and the transmon.
The flux qubit parameters $\Delta^{(p)}$, $g^{(pc)}_{\parallel}$ and $g^{(pt)}_{\parallel} \propto g^{(pc)}_{\parallel}$, strongly depend on the values of the offset-charge $n_g^{(p)} = C_gV_g/(2e)$ of the $\alpha$-junctions island and thus on the externally applied voltage gate $V_g$, as it can be seen in Fig.$\,$\ref{resonator_coupled_circuit_parameter_graphs}. In the figure, the values of the qubit frequency and the longitudinal coupling strength $g_\parallel \equiv g^{(pt)}_{\parallel}$ are plotted as a function of $n_g^{(p)}$ and for several values of $\alpha$ and the coupling capacitance $C_{cp}$. 
We notice that the longitudinal coupling vanishes when the qubit frequency is at its maximum, i.e. at $n_g^{(p)}=0$, as it can be understood from the Feynman-Hellmann theorem, giving $g_\parallel \propto \partial \Delta_p / \partial n_g^{(p)}$. Instead, when the coupling is at its maximum, i.e. at $n_g^{(p)}=\pm 1/2$, $\Delta_p$ vanishes. However, at any other value of $n_g^{(p)}$ the coupling can be conveniently tuned by both $\alpha$ and $C_{cp}$, as shown in Fig.$\,$\ref{resonator_coupled_circuit_parameter_graphs}(c)-(d).

We can show that Hamiltonian (\ref{eq:resonator_coupled_hamiltonian_eb0}) yields the required target-probe coupling, by applying the transformation $D(\hat\gamma) = e^{i\hat{\gamma} (\hat{a}^\dagger +\hat{a})}$, with $\hat{\gamma} = \left(g_\perp^{(ct)}\hat{n}^{(t)} +g_\parallel^{(pc)}\sigma_z^{(p)}\right)/\hbar\omega_r$.
In this way the transformed Hamiltonian $\tilde{H} = D(\hat\gamma)HD^\dagger(\hat\gamma)$ takes the form, up to constant terms:
\begin{align}
\label{transformed_hamiltonian}
        \hat{H} = &\, 4E_{Ct}(\hat{n}^{(t)})^2+E_{Jt}[1-\cos(\hat{\varphi}^{(t)})] + \frac{\Delta_p}{2} \hat{\sigma}^{(p)}_z + \hbar\omega_r  \hat{a}^\dagger \hat{a} \nonumber\\
        - &\frac{E_{Jt}}{2}\Big\{ (e^{i\tilde{\gamma}(\hat{a}^\dagger + \hat{a})}-1)e^{i\varphi_t} + (e^{-i\tilde{\gamma}(\hat{a}^\dagger + \hat{a})}-1)e^{-i\varphi_t} \Big\}
        \nonumber\\
        + & \frac{(g^{(ct)}_{\perp})^2}{\omega_r} \hat{n}_{t}^2 + \bigg(g^{(pt)}_{\parallel} +\frac{2g^{(ct)}_{\perp}g^{(pc)}_{\parallel}}{\omega_r} \bigg)\hat{\sigma}^{(p)}_z\hat{n}^{(t)} ,
\end{align}
with $\tilde{\gamma} = (g_\perp^{(ct)}/\hbar\omega_r)$. Since we are not working with a bosonic representation of the transmon Hamiltonian, the  $\sim D(\hat\gamma)\cos(\hat{\varphi}^{(t)})D^\dagger(\hat\gamma)$ term yields transmon-resonator interactions. 
In the tomography protocol described above, these terms act when the transmon is brought into the charge regime $E_{Jt}\sim0$, thus we can consider them negligible.
The term $(g^{(t)}_{\perp})^2 (\hat{n}^{(t)})^2/\hbar\omega_r$ contributes to an additional charging energy in the transmon. Finally, the last term in Eq.$\,$(\ref{transformed_hamiltonian}) represents the effective coupling $g_{tp} = g^{(pc)}_{\parallel}+ 2g^{(ct)}_{\perp}g^{(pc)}_{\parallel}/\omega_r$ and can be used for the tomographic reconstruction of the target system (the transmon) in the charge basis. 

\section{Conclusions}
In summary we have presented a novel protocol for implementing quantum state tomography of the ground state of a superconducting circuit comprising a generic weak link. Our method aims at reconstructing the density matrix in its charge basis. This can be seen as a complementary method to the study of the junction itself, and can provide insights into the specific physical platform. The method deploys concepts familiar to cavity state tomography to a system where the coupling term between the target and the probe, crucial for the required probabilities measurement, does not commute with the target system Hamiltonian. We also present and analyse a possible target-probe circuit where the required coupling is designed via a capacitive coupling to a flux qubit. The density matrix reconstruction is then performed via a non-linear least-square regression, with a system of equations derived from the analysis of the adiabatic evolution of the system. We demonstrate the validity of the method by applying it to the reconstruction of the ground state of the transmon. We find that the overlap between its wavefunctions at different values of $E_J$ is sufficient for the implementation of the least-square regression. Since the method strongly relies on the model used to construct the system of equations, it is self-consistent when tested against different models. This shows indeed that this tomographic protocol can validate physical models. 

\section*{acknowledgements}
This work is supported by funding from UKRI EPSRC grants EP/T001062/1, EP/L02263X/1 and the UK-Israel innovation researcher mobility scheme.

\appendix

\section{Circuit Analysis}
\label{Appendix_circuit_analysis}
In this appendix we analyse the proposed circuit shown in Fig. \ref{resonator_coupled_circuit}, and derive its Hamiltonian. We apply the method of nodes \cite{Rasmussen2021} and include the external flux of the probe qubit in the generalised flux of the $\alpha$-junction on the left of the loop. The Lagrangian takes the form:
\begin{align}
\label{Lagrangian}
        \mathcal{L} &{}= \frac{C_{Jp}}{2} ( \dot{\phi}_1^2 + \dot{\phi}_2^2 ) + \frac{\alpha C_{Jp}}{2} (\dot{\phi}_3 - \dot{\phi}_1)^2  \nonumber\\
        &{}+ \frac{\alpha C_{Jp}}{2} (\dot{\phi}_3 - \dot{\phi}_2)^2  + \frac{C_{g}}{2} (V_g - \dot{\phi}_3)^2 + \frac{C_r}{2}  \dot{\phi}^2_4 \nonumber\\
        &{} + \frac{C_{t,\Sigma}}{2} \dot{\phi}^2_5 + \frac{C_{cp}}{2} (\dot{\phi}_4 - \dot{\phi}_3)^2 + \frac{C_{ct}}{2} (\dot{\phi}_5 - \dot{\phi}_4)^2    \nonumber\\
        &{}- E_{Jp} \big[ 2(1 + \alpha) - \cos( 2\pi\phi_1) - \cos(2\pi\phi_2) \nonumber\\
        &{}- \alpha \cos( 2\pi(\phi_3 - \phi_1)) - \alpha \cos( 2\pi(\phi_2 - \phi_3 + f_p)) \big]\nonumber\\
        &{}- E_{Jt} [ 1 - \cos(2\pi\phi_5) ] - \frac{\phi_4^2}{2 L_r},
    \end{align}
where $C_{t,\Sigma} = (C_{Jt}+C_t)$ and we have set $\phi_0=1$. In the case of a symmetric split junction in the transmon, $E_{Jt}$ will depend on the external flux $f_t$ threading the junction loop. From Eq.$\,$(\ref{Lagrangian}) we can then calculate the node charges as the conjugate momenta associated with the node flux, $q_i = \partial \mathcal{L}/\partial \dot{\phi}_i$, i.e.
\begin{align}
    \label{eq:resonator_charge_equations}
        q_1 & = (1 + \alpha) C_{Jp} \dot{\phi}_1 - \alpha C_{Jp} \dot{\phi}_3 \nonumber\\
        q_2 & = (1 + \alpha) C_{Jp} \dot{\phi}_2 - \alpha C_{Jp} \dot{\phi}_3 \nonumber\\
        q_3 & \simeq -\alpha C_{Jp} ( \dot{\phi}_1 + \dot{\phi}_2) + (2 \alpha C_{Jp} + C_{cp}) \dot{\phi}_3 - C_{cp} \dot{\phi}_r \nonumber\\
        q_4 & = - C_{cp} \dot{\phi}_3 + ( C_{cp} + C_r + C_{ct} ) \dot{\phi}_4 - C_{ct} \dot{\phi}_5 \nonumber\\
        q_5 & = - C_{ct} \dot{\phi}_4 + ( C_{ct} + C_{t,\Sigma}) \dot{\phi}_5,
    \end{align}
where we have assumed $C_g \ll C_{Jp},C_{cp}$ and included the offset charge $C_gV_g$ in the definition of $q_3$, $q_3 \equiv \partial \mathcal{L}/\partial \dot{\phi}_3 +C_gV_g$.
These equations can be grouped into the capacitance matrix, defined by the relation $\mathbf{q} = \mathbf{C} \dot{\boldsymbol{\phi}}$, with $\mathbf{q} = [q_1,\dots, q_5]$ and $\boldsymbol{\phi}=[\phi_1,\dots,\phi_5]$, resulting in 
\begin{align}
\label{Cmat}
    &{}\mathbf{C} = \nonumber\\
    &{}\begin{pmatrix}
        \scriptstyle{(1 + \alpha) C_{Jp}} & \scriptstyle{}0 & \scriptstyle{-\alpha C_{Jp}} & \scriptstyle{0} & \scriptstyle{0} \\ 
        \scriptstyle{0} & \scriptstyle{(1 + \alpha) C_{Jp}} & \scriptstyle{-\alpha C_{Jp}} & \scriptstyle{0} & \scriptstyle{0} \\ 
        \scriptstyle{-\alpha C_{Jp}} & \scriptstyle{-\alpha C_{Jp}} & \scriptstyle{2 \alpha C_{Jp} + C_{cp}} & \scriptstyle{-C_{cp}} & \scriptstyle{0} \\ 
        \scriptstyle{0} & \scriptstyle{0} & \scriptstyle{-C_{cp}} & \scriptstyle{C_r + C_{cp} + C_{ct}} & \scriptstyle{-C_{ct}} \\
        \scriptstyle{0} & \scriptstyle{0} & \scriptstyle{0} & \scriptstyle{-C_{ct}} & \scriptstyle{C_{ct} + C_{t,\Sigma}}
    \end{pmatrix}.
\end{align}
By inverting Eq.$\,$(\ref{Cmat}), the Hamiltonian of the system can be found in terms of the charge and flux operators,
    $\mathcal{H} = \mathbf{q}^T \mathbf{C^{-1}} \mathbf{q} -\mathcal{L}(\mathbf{q},\mathbf{\Phi})$, leading to
\begin{align}
\label{hamcircuit}
    H &{}= \frac{\tilde{C}_{p,0}(2+\alpha)\alpha + \tilde{C}_{p,1}(1+\alpha)}{2(1 + \alpha)C_0^2}(q_1^2 + q_2^2) \nonumber\\
    &{}+\frac{\tilde{C}_{p,0}\alpha^2}{(1 + \alpha)C_0^2} q_1q_2 + \frac{\tilde{C}_{p,0}\alpha}{C_0^2} (q_1+q_2)q_3 \nonumber\\
    &{}+ \frac{\tilde{C}_{p,0}(1+\alpha)}{2C_0^2}q_3^2 +\frac{1}{2\tilde{C}_r} q_4^2 +\frac{1}{2\tilde{C}_t} q_5^2 \nonumber\\
    &{}+\frac{\tilde{C}_{cp}}{C_0^2}\left\{\alpha(q_1 + q_2) + (1+\alpha) q_3 \right\}q_4 \nonumber\\
    &{}+\frac{C_{cp}C_{ct}}{C_{Jp}C_0^2}\left\{\alpha(q_1 + q_2) + (1+\alpha) q_3 \right\}q_5  \nonumber\\
    &{}+\frac{C_{ct} \left( 2\alpha + C_{cp}(1+\alpha)/C_{Jp} \right)}{C_0^2} q_4 q_5   \nonumber\\
    &{}+E_{Jp} \Big[ 2(1 + \alpha) - \cos( 2\pi\phi_1) - \cos(2\pi\phi_2)] \nonumber\\
        &{}-\alpha \cos(2\pi( \phi_3 - \phi_1)) -\alpha\cos(2\pi( \phi_2 - \phi_3 + f_p)) \Big]\nonumber\\
        &{}+ E_{Jt} [ 1 - \cos(2\pi\phi_5) ] + \frac{\phi_4^2}{2 L_r},
\end{align}
where we have defined:
\begin{align}
    C_0^2 =&{}\, 2 \left[C_{t,\Sigma}(C_r + C_{ct} + C_{cp}) + C_{ct}(C_r + C_{cp})\right] \alpha\nonumber\\ 
    &{}+(C_{cp}/C_{Jp})\left[C_{t,\Sigma}(C_r + C_{ct}) + C_rC_{ct}\right] (1+\alpha); \\
    \tilde{C}_{p,0} = &{}\, \left[C_{t,\Sigma}(C_r + C_{cp} + C_{ct})  + C_{ct}(C_r + C_{cp})\right]/C_{Jp};\\
    \tilde{C}_{p,1} =&{}\, (C_{cp}/C_{Jp}^2)[(C_{ct} + C_r) C_{t,\Sigma} + C_{ct} C_r] ;\\
    \tilde{C}_{cp} = &{}\, C_{cp}(C_{ct}+C_{t,\Sigma})/C_{Jp} ;\\
    \tilde{C}_r = &{}\frac{C_{Jp} C_0^2}{(C_{t,\Sigma} + C_{ct})\left(2C_{Jp}\alpha + C_{cp}(1+\alpha)\right)};\\ 
    \tilde{C}_t =&{}\frac{C_{Jp} C_0^2}{2C_{Jp}(C_r + C_{cp}+C_{ct})\alpha + C_{cp}(C_{ct} + C_{r})(1+\alpha)}.
\end{align}
By re-expressing the terms containing $q_4^2$ and $q_5^2$ in the forms $q_4^2/2\tilde{C}_r$ and $q_5^2/2\tilde{C}_t$ respectively, we have obtained the renormalised capacitance of the resonator and the transmon, respectively. We can notice that the chosen circuit yields both a coupling term between the probe qubit and the resonator, $H_{pc}(q_1,q_2,q_3)\otimes q_4$, and a coupling term between the probe qubit and the transmon, $H_{pt}(q_1,q_2,q_3)\otimes q_5$. These differ by only a factor $H_{pt}/H_{pc} = C_{ct}/(C_{ct}+C_{t,\Sigma}) \sim C_{ct}/C_{t,\Sigma} \ll 1$. 
To analyse the couplings involved, it is convenient to project Hamiltonian (\ref{hamcircuit}) onto the subspace of the two lowest energy levels of the probe qubit, defining $\hat{\sigma}^{(p)}_k$ the Pauli operators related to the projected subspace. We also express the resonator's charge and phase degrees of freedom in terms of the bosonic operators, such that $\hat{n}_r = q_4/2e = -in_0(a^\dagger - a)$ and $\varphi_r = 2\pi\phi_r =\varphi_0(a^\dagger + a)$. The resulting Hamiltonian is, up to constant terms,
\begin{figure}[t!]
    \center    \includegraphics[width=0.8\linewidth]{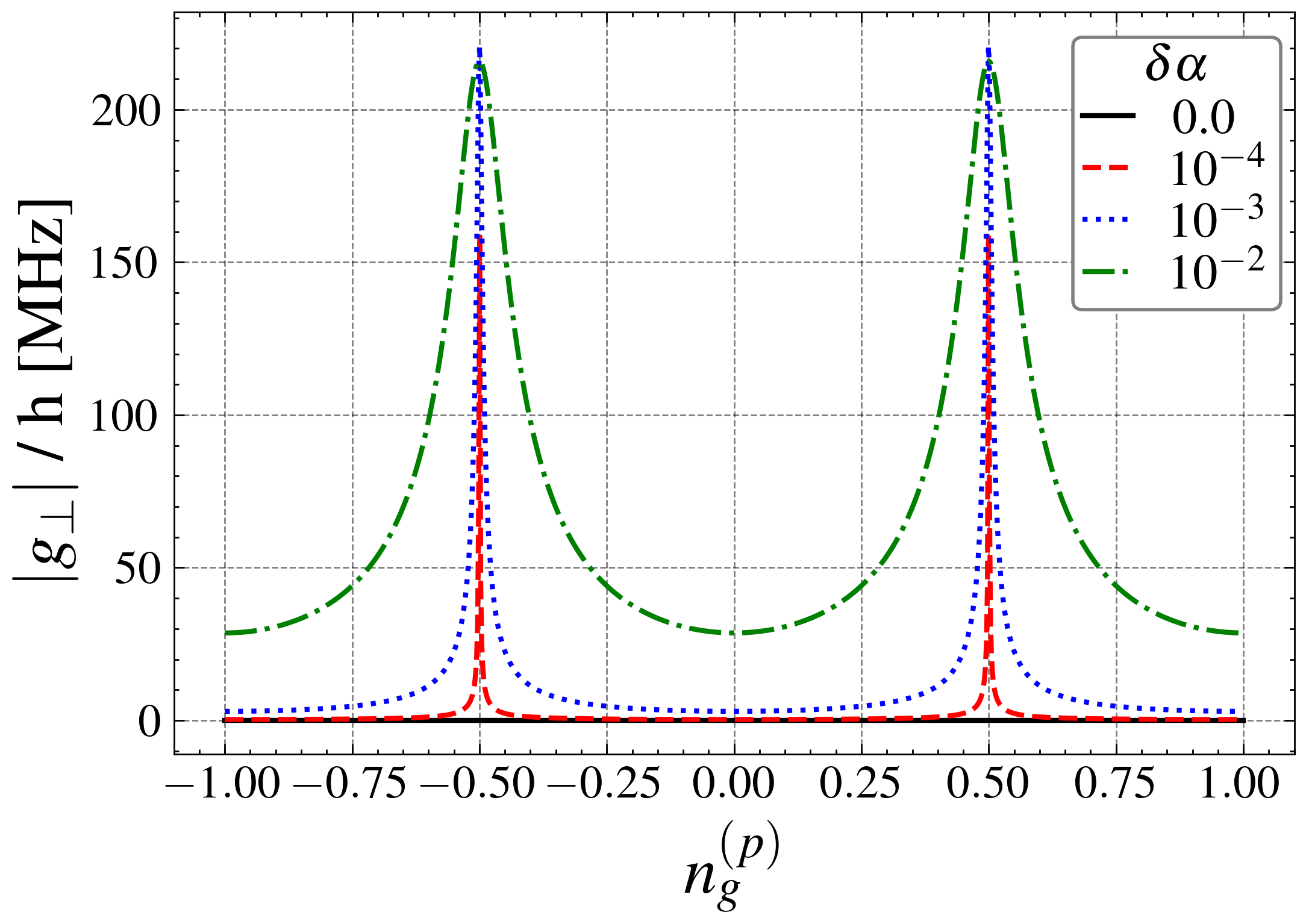}\hspace{10pt}
\caption{\label{delta_wrt_alpha_ng_alpha_dispersion} Transverse coupling between the probe and resonator, $g_{\perp}\equiv g_{\perp}^{(pc)}$, for the circuit represented in Fig. \ref{resonator_coupled_circuit}, with respect to the probe charge offset $n_g^{(p)} = q_3/(2e)$. The coupling is calculated for a range of discrepancies between the left/right junctions of the flux qubit $\alpha$-loop, i.e. varying $\delta\alpha = \alpha_R - \alpha_L$, where $\alpha_{L/R}$ indicates the Josephson energy reduction of the left/right branch, and $\alpha_L=0.4$. When $\delta\alpha \neq 0$, the transverse coupling assumes non-zero values, with a broadening spike centered at $n_g^{(p)}=\pm 1/2$. The other circuit parameters are set to: $f_p=0.5$, $E_{Jp}/h=121$ GHz, $E_{Jt}/h=5$ GHz, $C_{Jp}=8$ fF, $C_{Jt}=4$ fF, $C_t=40$ fF, $C_{cp}=5$ fF, $C_{ct}=5$ fF, $C_r=100$ fF, $L_r=10$ nH.
}
\end{figure}
\begin{align} \label{eq:resonator_coupled_hamiltonian_eb}
        \hat{H} = &\, 4E_{Ct}(\hat{n}^{(t)})^2-E_{Jt}[1-\cos(\hat{\varphi}^{(t)})] + \frac{\Delta_p}{2} \hat{\sigma}^{(p)}_z + \hbar\omega_r  \hat{a}^\dagger \hat{a} \nonumber\\
        - & i g^{(pc)}_{\parallel} \hat{\sigma}^{(p)}_z (\hat{a}^\dagger - \hat{a}) + g^{(pt)}_{\parallel} \hat{\sigma}^{(p)}_z \hat{n}^{(t)} \nonumber\\
        - & i \left(\mathrm{Re}\big[g^{(pc)}_{\perp}\big]\,\hat{\sigma}^{(p)}_x + i\,\mathrm{Im}\big[g^{(pc)}_{\perp}\big]\,\hat{\sigma}^{(p)}_y\right) (\hat{a}^\dagger - \hat{a}) \nonumber\\
        + &   \left(\mathrm{Re}\big[g^{(pt)}_{\perp}\big]\,\hat{\sigma}^{(p)}_x + i\,\mathrm{Im}\big[g^{(pt)}_{\perp}\big]\,\hat{\sigma}^{(p)}_y\right) \hat{n}^{(t)} \nonumber\\
        - & i g^{(ct)}_{\perp} (\hat{a}^\dagger - \hat{a}) \hat{n}^{(t)},
\end{align}
where $E_{Ct} = e^2/2\tilde{C}_t$ defines the renormalised charging energy of the transmon qubit, $\hat{n}^{(t)}=q_5/2e$ and $\hat{\varphi}^{(t)}=2\pi\phi_5$ are its charge and flux operators, $\Delta_p$ is the probe qubit energy gap, $g^{(pc)}_{\parallel}$ ($g^{(pc)}_{\perp}$) and $g^{(pt)}_{\parallel}$ ($g^{(pt)}_{\perp}$) are the resulting longitudinal (transverse) coupling strengths between the probe qubit and the resonator and the probe qubit and the transmon, respectively; $\omega_r = 1/2\sqrt{L_r\tilde{C}_r}$ is the renormalised frequency of the resonator, while $g^{(ct)}_{ \perp}$ is the transverse coupling between the resonator and the transmon. To determine the parameters in Eq.$\,$(\ref{eq:resonator_coupled_hamiltonian_eb}) one needs to diagonalise the probe qubit circuit part $\hat{H}_p(q_1,q_2,q_3)$ of Eq.$\,$(\ref{hamcircuit}) and evaluate the following quantities
\begin{align}
    \Delta_p &= \langle 1 | \hat{H}_p | 1 \rangle - \langle 0 | \hat{H}_p  | 0 \rangle; \\
    g^{(pc)}_\parallel &= \sqrt\frac{\hbar }{2Z_r}\langle -  | \hat{H}_{pc} | + \rangle; \\
    g^{(pc)}_\perp &= \sqrt\frac{\hbar }{2Z_r}\langle 0 | \hat{H}_{pc} | 1 \rangle;\\
    g^{(pt)}_\parallel &= (2e)\langle -  | \hat{H}_{pt} | + \rangle = \sqrt\frac{4e^2Z_r}{\hbar}\, g^{(pc)}_\parallel ;\\
    g^{(pt)}_\perp &= (2e)\langle 0 | \hat{H}_{pt} | 1 \rangle = \sqrt\frac{4e^2Z_r}{\hbar}\, g^{(pc)}_\perp ; \\
    g^{(ct)}_\perp &= \sqrt\frac{2\hbar e^2}{Z_r}\frac{2C_{ct} \left[ 2C_{Jp}\alpha + C_{cp}(1+\alpha) \right]}{C_{Jp}C_0^2},
\end{align}
where $Z_r = \sqrt{L_r/\tilde{C}_r}$, $| 0 \rangle,\,| 1 \rangle$ are the two lowest-energy probe qubit eigenstates, and $| \pm \rangle = (| 0 \rangle \pm | 1 \rangle)/\sqrt{2}$.

The numerical calculations of these couplings lead to $g_{\perp}^{(pc)}\to 0$ at the flux qubit sweet-spot $f_p=1/2$ and for a symmetric $\alpha$-loop, leaving a purely longitudinal coupling between the flux qubit and the cavity, and between the flux qubit and the transmon. For sake of completeness Figure \ref{delta_wrt_alpha_ng_alpha_dispersion} shows the effect of a discrepancy between the $\alpha$-parameters of the flux qubit junctions, $\delta\alpha = \alpha_R - \alpha_L$, on the transverse coupling $g_{\perp}\equiv g_{\perp}^{(pc)}$. The $\alpha_R$ parameter of the $\alpha$-loop's right branch was set to increasing values, showing a broadening spike in the transverse coupling, centered around $n_g^{(p)}=\pm 1/2$. Far from these points, however, there is room for a possible trade-off where, depending on the $\alpha$-loop asymmetry, the longitudinal coupling contribution can be higher that the transverse one. For the purpose of this paper, we keep the $\alpha$-loop symmetric and the flux qubit operating at its sweet-spot. This reduces Eq.$\,$(\ref{eq:resonator_coupled_hamiltonian_eb}) to Eq.$\,$(\ref{eq:resonator_coupled_hamiltonian_eb0}) used in the main text.
 
%\bibliography{tomography, references_quantum_tomography}

\end{document}